\documentclass{iopjournal}
\usepackage[T1]{fontenc}
\usepackage{lmodern}

\usepackage{amsmath}
\usepackage{tabularx}

\usepackage{braket}

\usepackage{ulem}
\begin{document}

\articletype{Paper} %	 e.g. Paper, Letter, Topical Review...

\title{Error signals for overcoming the laser power limits of gravitational-wave detectors}

\author{Liu Tao$^1$\orcid{0000-0003-4382-5507}, Pooyan~Goodarzi$^1$\orcid{0009-0008-1093-6706} and Jonathan W. Richardson$^{1,*}$\orcid{0000-0002-1472-4806}}

\affil{$^1$Department of Physics \& Astronomy, University of California, Riverside, Riverside, USA}

\email{jonathan.richardson@ucr.edu}


\begin{abstract}
A major barrier to improving the quantum-limited sensitivity of gravitational-wave observatories is the thermal distortions of the test masses which arise at megawatt laser power. Recent advances in a new form of higher-order wavefront correction, in which corrective heating profiles are applied to the test mass surfaces near their edges, together with other planned instrumental upgrades, have the potential to enable a tenfold reduction of the quantum noise floor of future detectors. However, realizing high levels of quantum noise reduction in practice hinges on identifying measurable error signals to finely control each wavefront actuator, in order to suppress wavefront errors to a few-nanometer precision across the full mirror apertures. No direct source of such an error signal exists in LIGO today. We demonstrate that thermally imaging the surface of each test mass, with a calibration provided by existing Hartmann wavefront sensors, can provide these critical error signals. We show that the surface temperature profiles obtained from thermal imaging can be uniquely mapped to a finite element model of the mirror whose complete thermal state is identified, enabling full-aperture wavefront reconstruction and direct error signals for real-time precision wavefront control. This new sensing capability can enable up to a 31\% strain sensitivity improvement in LIGO~A+ at 95\% confidence, increasing the sky-averaged detection range for binary neutron star mergers by 10~Mpc, and will be integral to a next-generation 40-km gravitational-wave observatory in the U.S., Cosmic Explorer.
\end{abstract}

\section{Introduction}
\label{sec:intro}

Since the first detection of gravitational waves from two coalescing black holes in 2015~\cite{GW150914}, the Laser Interferometer Gravitational-Wave Observatory (LIGO) and the European Virgo observatory have opened a new means of observing the Universe, launching an era of multi-messenger astronomy. A wide variety of merger events involving black holes and neutron stars have now been observed~\cite{GWTC-1, GWTC-2, GWTC-2.1, GWTC-3}. The path to opening a much broader range of observational science, from precision cosmology~\cite{Chen:2018, Farr:2019} to direct tests of strong-field general relativity~\cite{Skenderis:2008, Cardoso:2017, Brustein:2018} and dense nuclear matter~\cite{Tsang:2012}, now lies in reducing the instrumental noise floor of gravitational-wave detectors.

LIGO plans to incorporate technical improvements in two major upgrades of the 4-km detectors, LIGO~A+~\cite{Aplus, Aplus_Projection} followed by LIGO~$\rm A^{\#}$~\cite{PostO5Report:2022, LIGOwhitepaper2025}. The A+ upgrades are aimed primarily at reducing thermal noise through improved optical coatings. The $\rm A^{\#}$ upgrades will include larger 100-kg test masses, new improved suspensions, and significantly higher levels of circulating laser power and quantum squeezing~\cite{PhysRevD.104.062006, PhysRevX.13.041021}. Together, these upgrades will demonstrate the key technology for a next-generation 40-km gravitational-wave observatory in the U.S., Cosmic Explorer~\cite{CEHorizonStudy}. With 320-kg test masses and a tenfold longer arm length, Cosmic Explorer will push the gravitational-wave detection horizon to near the edge of the observable Universe, detecting hundreds of thousands to millions of compact binary merger events per year. 

Quantum noise, arising from the quantum nature of the laser field~\cite{Caves:1980, Caves:1981}, is the limiting source of instrumental noise across most of the frequency band accessible to ground-based gravitational-wave detectors. Amplitude-quadrature fluctuations of the laser field displace the test masses through radiation pressure, introducing significant noise at low frequencies (below 20~Hz). Phase-quadrature fluctuations of the laser field also manifest as shot noise in the detector's phase-sensitive readout signal. Photon shot noise limits LIGO's strain sensitivity at all frequencies above 200~Hz. In order to reach their strain sensitivity targets, future detectors must achieve significantly greater levels of quantum noise reduction, through higher laser power and squeezing. The LIGO~A+ detectors will require 750~KW of circulating laser arm power, roughly double the arm power achieved in Advanced LIGO~\cite{o4_instrumentation}, and 7~dB of frequency-dependent squeezing~\cite{Aplus_Projection}. Following A+, the LIGO~$\rm A^{\#}$ and Cosmic Explorer detectors must achieve an unprecedented 1.5~MW of circulating laser arm power and 10~dB of frequency-dependent squeezing~\cite{PostO5Report:2022, CEHorizonStudy}.

The achievable arm power and squeezing are limited, in practice, by thermal distortions of the test masses. Thermal distortions arise from the absorption of laser power by the mirror coatings. Roughly 0.5~ppm of uniform coating absorption is expected in future LIGO optics. In addition, small coating contaminants known as ``point absorbers'' were a major source of non-uniform absorption in Advanced LIGO's O2 and O3 observing runs~\cite{Brooks:2021, Jia:2021}, although LIGO has made significant progress towards eliminating them from its O4 optics. Absorption creates temperature gradients within the test masses that drive thermo-elastic deformation of the reflective surfaces and thermo-refractive lensing in the substrates~\cite{Brooks:2016}. The magnitudes of these aberrations increase proportionally to the incident laser power. They imprint wavefront errors on the circulating optical fields that result in laser power loss, squeezing loss, and mis-rotation of the squeezing angle~\cite{McCuller:2021}, directly increasing the quantum noise floor of the detector. Modeling indicates that overcoming thermal distortions to achieve the high levels of laser power and squeezing required by LIGO~$\rm A^{\#}$ and Cosmic Explorer will require significant advances in wavefront error sensing and correction~\cite{AsharpTCSReqs:2022}.

In recent work~\cite{FROSTI_PRL, Tyler_2025_Optica}, we reported the potential of a new class wavefront actuator, known as the FROnt Surface Type Irradiator (FROSTI), to enable next-generation quantum noise targets when used to augment LIGO's existing thermal compensation capabilities~\cite{Brooks:2016}. The FROSTI projects a corrective annular heating pattern onto the front surface of each test mass near the edge. However, \textit{realizing} the potential sensitivity improvements of Ref.~\cite{FROSTI_PRL} relies on there also existing error signals to precisely identify the optimal power settings of the wavefront actuators, so as to minimize root mean square (RMS) wavefront errors across the full aperture of the mirrors. No direct error signals containing full-aperture wavefront information exist in LIGO today. LIGO's existing Hartmann wavefront sensors~\cite{Brooks:2016} are fundamentally limited in their field of view by the size of the probe beam. For optical systems with large optics, such as gravitational-wave detectors, this restriction constrains the field of view to only the central region of the optics. Similarly, Virgo employs phase cameras as wavefront sensors to measure the spatial structure of the laser fields, but these systems also suffer from a limited field of view due to the limited beam size~\cite{van_der_Schaaf:2016}.

In this paper, we present a novel sensing technique that can provide these critical error signals. The technique uses thermal imaging cameras to effectively expand the field of view of LIGO's existing Hartmann wavefront sensors, obtaining sensitivity to wavefront errors across the full aperture of the mirrors. It was directly inspired by experimental results obtained from FROSTI prototype testing~\cite{Tyler_2025_Optica}. During this testing, the FROSTI illuminated a 40~kg LIGO test mass in vacuum while the optic’s response was observed with both a thermal imaging camera (viewing the entire mirror surface) and a Hartmann wavefront sensor (viewing a smaller area, due to its limited probe beam size). One of the key findings was that independently fitting the thermal imaging measurement to a finite element analysis (FEA) model of the test mass could accurately reproduce the Hartmann wavefront sensor measurement~\cite{Tyler_2025_Optica}. It demonstrated that the surface temperature profile is an accurate predictor of the thermo-elastic and thermo-refractive wavefront distortions within the bulk optic, at least when the FROSTI is the only thermal source heating the test mass. 

This experimental observation motivated us to investigate further the possibility of using the same FEA-model-referred technique in a more realistic case with multiple heating sources acting on the test mass. Our work leverages simulations to explore the extent to which this unique mapping can be recovered from thermal images when laser beam heating, ring heater thermal actuation, and FROSTI thermal actuation are \textit{all} present simultaneously. We find that that the thermal state of the test masses can still be uniquely and accurately recovered from thermal images when the optics are in a thermal steady state. This condition will be ensured in future LIGO operation, as the A+ upgrades will include an offline 10.6~$\mu \rm m$ laser heating system for each input test mass (ITM), known as the Central HEater for Transient Attenuation (CHETA)~\cite{CHETA_concept}, designed to maintain the thermal steady state of the optic any time the main laser beam is not present. The proposed thermal imaging cameras can, in fact, be used to precisely match the size, location, and power of the CHETA beam to the main laser beam, improving its effectiveness at thermal transient suppression.

In practice, LIGO's Hartmann wavefront sensors are still needed to accurately calibrate the raw fluxes measured by the thermal imaging cameras to units of surface temperature. This calibration depends on several parameters that are difficult to measure in situ, including the transmissivity of the vacuum viewport, the effective emissivity of the test mass coating, and the viewing distance. The raw surface temperature map is calibrated by requiring it to self-consistently agree with the actual level of wavefront error measured by the Hartmann wavefront sensor in the central region of the test mass, where there is overlapping coverage by both sensors. The two quantities are related to each other using an FEA model of the test mass. The calibrated surface temperature map then provides thermal state information across the entire aperture of the optic, including the FROSTI actuation which lies outside the field of view of the Hartmann wavefront sensors. This allows the full-aperture wavefront phase maps to be reconstructed from the FEA model, equivalently as if the Hartmann wavefront sensor’s probe beam could be expanded to cover the full surface of the optic.

We describe our thermal-imaging-enhanced wavefront error sensing technique in \S\ref{sec:procedure} and demonstrate its feasibility in \S\ref{sec:performance}, assuming realistic sensor resolution limits. We then show in \S\ref{sec:impact} that the improved capability provided by this technique to infer, and thus to more finely correct, thermally-driven wavefront errors in gravitational-wave detectors will significantly improve their quantum-limited strain sensitivity, which we find can be improved up to 31\% in LIGO~A+. Finally, we discuss the impact of this work on future gravitational-wave detectors, beyond A+, in \S\ref{sec:conclusion}.

\section{Thermal State Inference}
\label{sec:procedure}

We aim to uniquely and accurately infer the full thermal state of each test mass directly from the temperature profile of its front (reflective) surface. This surface temperature map is measured in situ using a thermal imaging camera, located outside the vacuum system, and matched to an FEA model of the test mass. In this work, we do not make any specific assumption about the placement of the thermal camera or the design of its imaging optics, as the optical design is likely to be highly specific to a particular gravitational-wave detector based on the availability of vacuum viewports and the locations of baffles and other mechanical obstructions. However, any optical design must meet the general spatial and temperature resolution requirements identified in section~\ref{sec:performance}. This will generally favor camera placements within several meters of the test mass, for maximum numerical aperture, facing its front surface and with an unobstructed view of the entire surface.

Our FEA model treats a 40~kg test mass in vacuum, following the methodology of Ref.~\cite{Tyler_2025_Optica}. We consider the test mass in isolation, without the surrounding suspension cage, reaction chain, and other detector-specific hardware, but our model can be easily extended to include such detail for a detector-specific design analysis. The stationary heat transfer solution is solved assuming constant inward heat flux from each thermal source and radiative cooling with an effective surface emissivity of 0.90. From the steady-state temperature gradient in the optic, the thermo-elastic deformation of the reflective surface and bulk thermo-refractive lens are both calculated. Convergence analysis indicates that, for fused silica optics, a grid resolution of approximately 5~mm is required. The FEA model is parameterized by a set of thermal ``state'' parameters representing the heating powers of the thermal sources and the position of the laser beam on the optic's surface, which are fitted to best reproduce the measured surface temperature map. The best-fit FEA model of the test mass can then be used to produce full-aperture mirror phase maps representing the thermally induced wavefront errors arising in reflection and transmission, as uniquely determined by its thermal state.

\begin{figure}[t]
    \centering
    \includegraphics[width=0.65\linewidth]{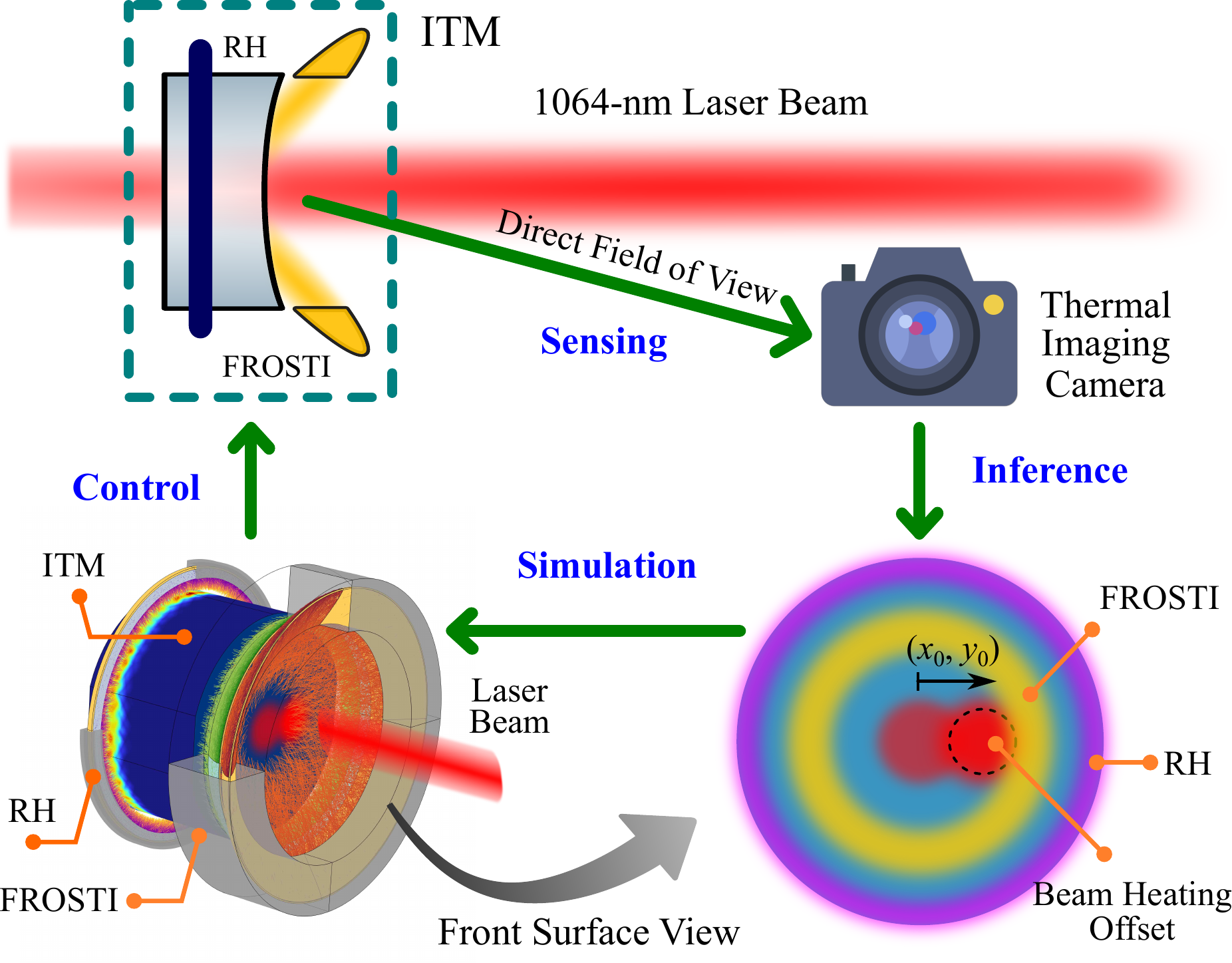}
    \caption{Test mass thermal state inference from front-surface thermal imaging. This procedure is used to infer the parameters in table~\ref{tab:parameters}.}
    \label{fig:decoder_concept}
\end{figure}

As illustrated in figure~\ref{fig:decoder_concept}, the thermal steady state of a test mass is fully characterized by five parameters: the power absorbed from the 1064~nm laser beam, $P_{\rm S}$; the beam's location on the test mass, $(x_0, y_0)$, which in general can be offset from the center; and the absorbed powers delivered by two different wavefront actuators. The actuators include LIGO's existing ring heater~\cite{Brooks:16}, which projects a heating power of $P_{\rm R}$ onto the barrel of the test mass to correct defocus (radius of curvature) error, and the recently proposed FROnt Surface Type Irradiator (FROSTI)~\cite{FROSTI_PRL, Tyler_2025_Optica}, which projects an annular heating pattern of power $P_{\rm F}$ onto the front surface of the test mass, near the edge, to correct higher-order aberrations. These parameters are summarized in table~\ref{tab:parameters}. The LIGO detectors also contain transmissive compensation plates located in the recycling cavity path just outside each arm cavity~\cite{Brooks:2016}. Each compensation plate is illuminated by a 10.6-$\mu \rm m$ laser heating beam whose intensity profile is reshaped via a beam mask. Unlike the CHETA system, which will be used offline as a surrogate for the main laser beam heating, the compensation plates are used as online actuators to partially mitigate the thermal lensing of the ITMs. While we restrict our analysis to only the test masses, the same procedure could also be applied to the compensation plates to sense their full-aperture wavefront actuation in transmission.

In the steady state, the two-dimensional surface temperature increase of the test mass (above ambient temperature) measured by the thermal imaging camera can be decomposed into a linear combination of contributions from four thermal sources:
\begin{equation}
T(x, y) = P_{\rm S} \, \hat{T}_{{\rm S}}(x, y \,|\, x_0, y_0) + P_{\rm R} \, \hat{T}_{\rm R}(x, y) + P_{\rm F} \, \hat{T}_{\rm F}(x, y) +\delta P \, \hat{T}_{\rm \delta}(x, y)
\label{eq:total_temp}
\end{equation}
The functions $\hat{T}_{\rm R}(x,y)$ and $\hat{T}_{\rm F}(x,y)$ respectively represent the surface temperature maps due to the ring heater and FROSTI actuators delivering unit power. Similarly, the function $\hat{T}_{{\rm S}}(x, y \,|\, x_0, y_0)$ represents the surface temperature map due to unit power absorbed from the incident laser beam, whose centroid is located at position $(x_0, y_0)$ relative to the center of the test mass. The function $\hat{T}_{\rm \delta}(x, y)$ represents the surface temperature map due to unit power absorbed from all other environmental effects. In practice, we expect the power absorbed from the surrounding environment, $\delta P$, to be small relative to the powers delivered by the three active heating sources, $P_{\rm S}$, $P_{\rm R}$, and $P_{\rm F}$. We thus assume that the $\delta P$ term is zero in this study, but we discuss its practical implications further in section~\ref{sec:conclusion}. We refer to $\hat{T}_{\rm S}$, $\hat{T}_{\rm R}$, and $\hat{T}_{\rm F}$ as unit temperature maps or ``unit maps.'' The linear addition of units maps as assumed in equation~\ref{eq:total_temp} is confirmed by FEA modeling to be valid up to and beyond the power levels of interest ($P_{\rm S} \sim 1\;{\rm W}$ and $P_{\rm R} \sim P_{\rm F} \sim 10\;{\rm W}$).

We can infer the values of the three absorbed powers by computing the spatial overlap integrals of the measured temperature map with each unit map. For succinctness, we adopt a bra-ket notation to represent the spatial overlap integral of any two temperature maps, $T_{\rm \alpha}$ and $T_{\rm \beta}$, as
\begin{equation}
\braket{T_{\rm \alpha} | T_{\rm \beta}} = \iint T_{\rm \alpha}(x, y) \, T_{\rm \beta}(x, y) \, dx\,dy
\label{eq:overlap_integral2}
\end{equation}
evaluated over the entire surface of the test mass. The spatial overlap integral of the measured temperature map (equation~\ref{eq:total_temp}) with any unit map, $\hat{T}_{\rm \alpha}$, can thus be expressed as
\begin{equation}
\braket{\hat{T}_{\rm \alpha} | T} = P_{\rm S} \braket{\hat{T}_{\rm \alpha} | \hat{T}_{\rm S}} + P_{\rm R} \braket{\hat{T}_{\rm \alpha} | \hat{T}_{\rm R}} + P_{\rm F} \braket{\hat{T}_{\rm \alpha} | \hat{T}_{\rm F}}
\label{eq:overlap_integral}
\end{equation}
where the explicit dependence of $\hat{T}_{\rm S}$ on the laser beam's position, $(x_0, y_0)$, is dropped but remains implied.

\begin{table}[t]
    \centering
    \begin{tabularx}{\columnwidth}{c c l}
    \hline \hline 
    \textbf{Parameter} & \textbf{Units} & \textbf{Description}\\
    \hline
    $P_{\rm S}$ & W & Power absorbed from laser beam \\
    $x_{0}$ & mm & Laser beam centroid $x$-coordinate\\
    $y_{0}$ & mm & Laser beam centroid $y$-coordinate \\
    $P_{\rm R}$ & W & Power delivered by ring heater \\
    $P_{\rm F}$ & W & Power delivered by FROSTI \\
    \hline \hline
    \end{tabularx}
    \caption{Parameters defining the thermal steady state of one test mass. Each parameter is described further in \S\ref{sec:procedure}.}
    \label{tab:parameters}
\end{table}

Evaluating equation~\ref{eq:overlap_integral} for all three unit maps yields the matrix equation
\begin{equation}
\begin{pmatrix}
\braket{\hat{T}_{\rm S} | T} \\
\braket{\hat{T}_{\rm R} | T} \\
\braket{\hat{T}_{\rm F} | T} 
\end{pmatrix}
=
\begin{pmatrix}
\braket{\hat{T}_{\rm S} | \hat{T}_{\rm S}} & \braket{\hat{T}_{\rm S} | \hat{T}_{\rm R}} & \braket{\hat{T}_{\rm S} | \hat{T}_{\rm F}} \\
\braket{\hat{T}_{\rm R} | \hat{T}_{\rm S}} & \braket{\hat{T}_{\rm R} | \hat{T}_{\rm R}} & \braket{\hat{T}_{\rm R} | \hat{T}_{\rm F}} \\
\braket{\hat{T}_{\rm F} | \hat{T}_{\rm S}} & \braket{\hat{T}_{\rm F} | \hat{T}_{\rm R}} & \braket{\hat{T}_{\rm F} | \hat{T}_{\rm F}}
\end{pmatrix}
\begin{pmatrix}
P_{\rm S} \\
P_{\rm R} \\
P_{\rm F}
\end{pmatrix}
\end{equation}
which can be inverted to obtain an expression for the absorbed powers from the three thermal sources:
\begin{equation}
\begin{pmatrix}
P_{\rm S} \\
P_{\rm R} \\
P_{\rm F}
\end{pmatrix}
=
\begin{pmatrix}
\braket{\hat{T}_{\rm S} | \hat{T}_{\rm S}} & \braket{T_{\rm S} | \hat{T}_{\rm R}} & \braket{\hat{T}_{\rm S} | \hat{T}_{\rm F}} \\
\braket{\hat{T}_{\rm R} | \hat{T}_{\rm S}} & \braket{\hat{T}_{\rm R} | \hat{T}_{\rm R}} & \braket{\hat{T}_{\rm R} | \hat{T}_{\rm F}} \\
\braket{\hat{T}_{\rm F} | \hat{T}_{\rm S}} & \braket{\hat{T}_{\rm F} | \hat{T}_{\rm R}} & \braket{\hat{T}_{\rm F} | \hat{T}_{\rm F}}
\end{pmatrix}^{-1}
\begin{pmatrix}
\braket{\hat{T}_{\rm S} | T} \\
\braket{\hat{T}_{\rm R} | T} \\
\braket{\hat{T}_{\rm F} | T} 
\end{pmatrix}
\label{eq-decompose}
\end{equation}
An FEA thermal model of the test mass is used to calculate each unit map, from which the $3 \times 3$ matrix of overlap coefficients in equation~\ref{eq-decompose} can be numerically calculated.

Although the beam heating unit map in equation~\ref{eq-decompose}, $\hat{T}_{{\rm S}}$, depends on the beam's unknown position on the test mass, the absorbed powers and the beam position can be jointly inferred via an iterative procedure. First, we solve equation~\ref{eq-decompose} assuming that the beam is perfectly centered, using the unit map $\hat{T}_{{\rm S}}(x, y \,|\, 0,0)$. Second, using these initial estimates of the absorbed powers, we subtract the ring heater and FROSTI contributions from the measured temperature map to obtain a residual map representing the beam heating. We estimate the \textit{true} position of the beam's centroid as the location of the global maximum in this residual temperature map. Third, we solve equation~\ref{eq-decompose} again using the beam heating unit map for the estimated centroid, $\hat{T}_{{\rm S}}(x, y \,|\, x_0, y_0)$. Thus, we obtain an estimate of the actual beam position on the test mass, $(x_0, y_0)$, from step~2 and then use this to refine our estimate of the absorbed powers, $P_{\rm S}$, $P_{\rm R}$, and $P_{\rm F}$, in step~3.

With knowledge of the three absorbed powers and the laser beam position, direct error signals for the two wavefront actuators can be constructed from the difference between the inferred actuator powers and their optimal values for the inferred level of absorbed laser power ($P_{\rm R,0}$ and $P_{\rm F,0}$ for the ring heater and FROSTI, respectively):
\begin{align}
\varepsilon_{\rm R} &= P_{\rm R} - P_{\rm R, 0} \\
\varepsilon_{\rm F} &= P_{\rm F} - P_{\rm F, 0}
\label{eq:errorsignal}
\end{align}
Since the optimal actuator powers scale linearly with the absorbed laser power (i.e., the \textit{ratios} of optimal actuator powers to self-heating power are invariant), these error signals optimize the ring heater and FROSTI power settings for each test mass with minimal computational cost, without requiring running expensive FEA models on the fly. Minimizing these error signals thus minimizes the actual wavefront errors in reflection and transmission across the full mirror aperture.

\section{Model Performance}
\label{sec:performance}

To assess the inferential power of our model, we trial it on a set of simulated thermal images generated by an FEA model of a test mass with all three active heating sources. We assume fixed absorbed powers similar to those expected at 1.5~MW of arm power in LIGO~$\rm A^{\#}$: $P_{\rm S} = 1~{\rm W}$ of power is absorbed by the front surface of the test mass from the incident laser beam, $P_{\rm R} = 20~{\rm W}$ of power is absorbed by the barrel of the test mass from a ring heater heating profile matching the current LIGO design~\cite{Brooks:16}, and $P_{\rm F} = 10~{\rm W}$ of power is absorbed by the front surface of the test mass from a FROSTI heating profile matching the O5 prototype design~\cite{FROSTI_Optica}. The incident laser beam is radially offset from the test mass center by an increasing amount up to 20~mm. We generate the same thermal images at a number of different spatial and temperature resolutions. The spatial resolution represents the pixelation scale of the thermal image. The temperature resolution represents the minimum detectable temperature difference between pixels, which is imposed by the total readout noise of each microbolometer in the camera's sensor array.

\begin{figure}[htbp]
    \centering
    \includegraphics[width=1\linewidth]{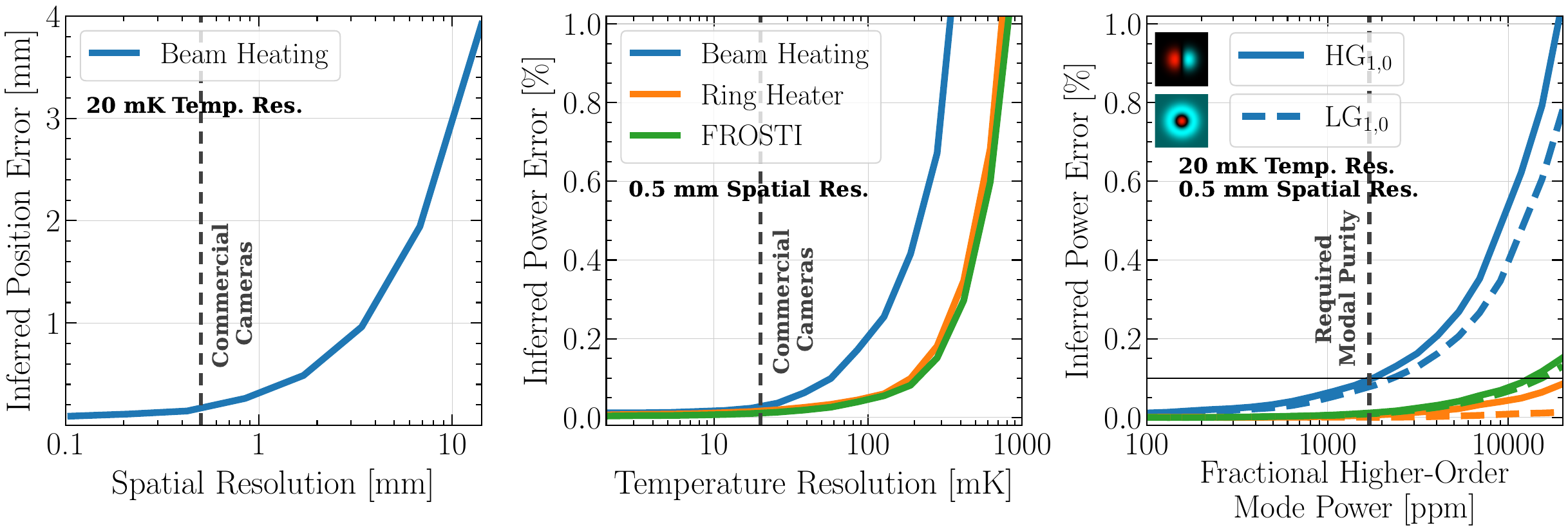}
    \caption{Performance of the thermal state inference algorithm. \textit{Left:} Error of the inferred radial position of the laser beam, as a function of the spatial resolution of the thermal image. \textit{Middle:} Relative errors of the inferred powers, as a function of the temperature resolution of the thermal image. \textit{Right:} Relative power errors versus the fractional higher-order mode power, shown for the first-order HG$_{1,0}$ mode (solid lines) and second-order LG$_{1,0}$ mode (dashed lines). These modes arise in laser cavities from small misalignments and mode-mismatches, respectively~\cite{Anderson:84}.}
    \label{fig:pos_prediction_error_num_pts}
\end{figure}

Figure~\ref{fig:pos_prediction_error_num_pts} shows the model's ability to accurately recover the true thermal state of the test mass, as the spatial and temperature resolutions of the thermal images are varied. The left panel shows the average error on the inferred radial position of the laser beam as a function of the spatial resolution (pixelation scale) of the image. The middle panel shows the average relative errors on the inferred power levels of the contributing thermal sources as a function of the temperature resolution (sensitivity) of the image. We find that a typical commercial thermal imaging camera with a $600 \times 600$ sensor array and a sensitivity of 20~mK (indicated by the dashed vertical lines in figure~\ref{fig:pos_prediction_error_num_pts}), assuming an appropriate set of image relay lenses, can accurately recover the true beam position to better than 0.5~mm and the true power levels to better than 0.1\% for all thermal sources. The feasibility of similar thermal imaging systems has already been demonstrated in Virgo to monitor for the presence of point absorbers on the test masses~\cite{VirgoLogbookPA}. 

The thermal state model assumes that, in heating the test mass surface, the primary laser beam has the intensity profile of a pure fundamental Gaussian mode. The presence of higher-order spatial mode content can introduce systematic error that degrades the predictive accuracy of the model. We quantify this error for the two lowest-order sources of higher-order mode content in the LIGO arm cavities. As shown by Ref.~\cite{Anderson:84}, a small angular or translational misalignment of the injected laser beam's axis relative to the cavity's geometrical axis results in a coupling of power into the first-order Hermite-Gaussian modes of the cavity, HG$_{1,0}$ and HG$_{0,1}$. Likewise, a small mode-mismatch of the injected laser beam (a mismatch of the beam's waist size and/or location) to the cavity's resonant spatial mode results in a coupling of power into the second-order Laguerre-Gaussian mode of the cavity, LG$_{1,0}$. Absent a high-order aberration such as a point absorber (which we do not assume in this study, but whose practical implications we discuss in section~\ref{sec:conclusion}), the HG$_{1,0}$, HG$_{0,1}$, and LG$_{1,0}$ modes are expected to constitute the dominant higher-order mode content of the arm cavities. The right panel of figure~\ref{fig:pos_prediction_error_num_pts} shows the effect of these higher-order modes on the inferred power error of our model. To achieve an inferential accuracy of 0.1\%, we find that the fraction of power in these higher-order modes must not exceed 1700~ppm, which is readily achievable. For comparison, this requirement is 13~orders of magnitude higher than recent estimates of the fractional power of first-order modes in the LIGO arm cavities~\cite{aLOG_HOM_content}. While other higher-order modes will also contribute to the total higher-order mode content, they are expected to be of similar magnitude or smaller than the first-order mode contribution.

With thermal imaging of the test masses, we estimate that, given a true level of laser beam heating, the $3\sigma$ inferential uncertainties in determining the optimal beam position and optimal ring heater and FROSTI power settings reduce to approximately 0.5~mm and 0.1\%, respectively, based on the findings in figure~\ref{fig:pos_prediction_error_num_pts}. This is in comparison to the current scenario without thermal imaging, in which we assume the same inference is performed using the estimated beam centroids on the test masses (from infrared camera feeds), the estimated arm cavity powers (from transmission monitor signals), and ex situ measurements of the test masses' uniform coating absorptivities. In this case, we estimate a $3\sigma$ inferential uncertainty of 1~mm on the beam position, based on the current best knowledge provided by infrared cameras used to monitor the beam position on the test masses~\cite{aLOG}. We estimate the laser power absorbed by each test mass as the product of the incident arm power, with average calibration uncertainty of 3.1\%~\cite{o4_instrumentation}, and the absorptivity of the reflective coating, with an average measurement uncertainty of 31.4\%~\cite{LIGOTMCoating}. This yields an overall uncertainty of 35.5\% in the power absorbed by each test mass, which represents the $3\sigma$ inferential uncertainty in determining the optimal ring heater and FROSTI power settings. We do not include Hartmann wavefront sensor signals in these uncertainty estimates because they have little sensitivity to the FROSTI actuators (which actuate outside their limited field of view) and because their fields of view (defined by the alignment of their probe beams) are known to drift with detector realignments, potentially introducing systematic error. Our uncertainty estimates are summarized in table~\ref{tab:uncertainties}.

\begin{table}[h]
    \centering
    \begin{tabularx}{\columnwidth}{l c c}
    \hline \hline 
    \textbf{Inferential Uncertainty ($\mathbf{3}\boldsymbol{\sigma}$)} & \textbf{With Thermal Imaging} & \textbf{Without Thermal Imaging}\\
    \hline
    Optimal beam $x$-position & 0.5~mm & 1.0~mm \\
    Optimal beam $y$-position & 0.5~mm & 1.0~mm \\
    Optimal ring heater power setting & 0.1\% & 35.5\% \\
    Optimal FROSTI power setting & 0.1\% & 35.5\% \\
    \hline \hline
    \end{tabularx}
    \caption{Inferential uncertainties in determining the optimal beam positioning and optimal wavefront actuator power settings for each test mass, estimated with and without the proposed thermal imaging cameras (see section~\ref{sec:performance}).}
    \label{tab:uncertainties}
\end{table}

\begin{figure}[h]
    \centering
    \includegraphics[width=1\linewidth]{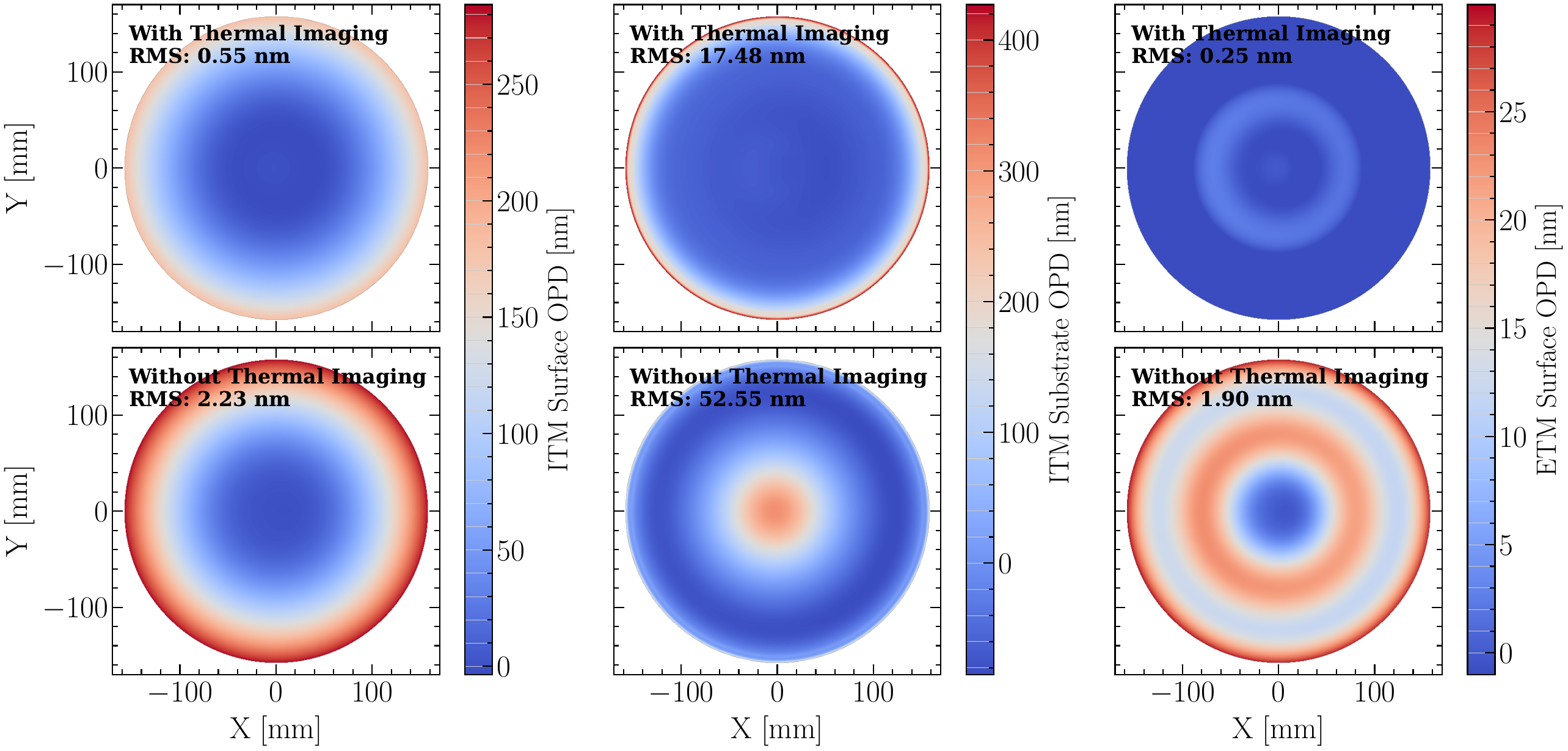}
    \caption{Residual wavefront error phase maps for the ITM surface, ITM substrate, and ETM surface (left to right), in units of optical path difference (OPD). The top and bottom rows show results with and without thermal-imaging-based sensing, respectively, at 1~W of absorbed laser power and the maximum $3\sigma$ power and position errors listed in table~\ref{tab:uncertainties}. The beam intensity profile-weighted RMS error for each map is also indicated.}
    \label{fig:error_signal_phase_maps}
\end{figure}

Figure~\ref{fig:error_signal_phase_maps} shows an example of the residual wavefront errors that could arise under these levels of inferential uncertainty, at their most extreme. The three columns, from left to right, correspond to the residual wavefront errors for the ITM surface, ITM substrate, and end test mass (ETM) surface at 1~W of absorbed laser power. The two rows, from top to bottom, correspond to the two scenarios with and without front-surface thermal imaging. For each panel, the residual RMS wavefront error of the mirror map is calculated, weighted by the Gaussian intensity profile of the main laser beam. The residual RMS wavefront error is smaller by a factor of 3-7 in the maps with thermal-imaging-based sensing, compared to the corresponding cases without it, demonstrating a significant improvement in the capability to correct thermally-driven wavefront distortions in both reflection and transmission.

\section{Astrophysical Impact}
\label{sec:impact}
Enhancing our ability to infer, and thus to more finely correct, thermally-driven wavefront errors in LIGO will directly improve its astrophysical sensitivity. To quantify its impact, we perform head-to-head Monte Carlo optical simulations of a LIGO~A+ detector with and without the proposed new sensing capabilities using \textsc{Finesse}~\cite{FinesseSoftware}. We integrate FEA thermal models of each test mass into the simulation to apply mirror phase maps due to laser beam heating, ring heater thermal actuation, FROSTI thermal actuation, and compensation plate thermal actuation. The laser beam heating is modeled assuming a uniform coating absorptivity of 0.5~ppm per test mass. We assume that there are no point absorbers near the center of the test masses and that larger scale non-uniformities are negligible, consistent with the properties of LIGO's O4 optics. We adopt the recommended FROSTI heating profiles from Ref.~\cite{FROSTI_PRL} and an ideal inverse-heating profile for the compensation plates (the inverse of the Gaussian intensity profile of the main laser beam).

To self-consistently solve the coupled thermal and optical problem, we follow the procedure of Ref.~\cite{FROSTI_PRL} with a generalization to allow for asymmetrical aberrations between the two arm cavities. Under this general approach, we first obtain FEA-generated mirror maps of the test masses for a target arm power, with optimized amounts of thermal actuator power applied, and then calculate the required input power from the optical gains of the arm cavities and power recycling cavity. This inverted approach of treating the arm power as the \textit{independent} variable, and the input power as the dependent variable, enforces consistency between the incident power levels assumed in the FEA simulations generating the mirror maps and the optical simulations then loading those maps. However, Ref.~\cite{FROSTI_PRL} treats only identical thermally-driven aberrations in the two arm cavities, whereas our present Monte Carlo approach introduces \textit{independent} random errors in each arm, leading to differential mode-mismatch and arm power imbalances of up to several percent. To allow for this, we generalize the procedure of Ref.~\cite{FROSTI_PRL} to the following iterative approach:
\begin{enumerate}
    \item Take as the independent variable the average arm cavity power, $P_{\rm avg}$.
    \item Apply mirror maps to the test masses initially assuming that the incident power on each is $P_{\rm avg}$.
    \item Run the \textsc{Finesse} simulation to obtain initial estimates of the $x$-arm, $y$-arm, and power recycling cavity gain factors, $G_{\rm X}$, $G_{\rm Y}$, and $G_{\rm PR}$, respectively.
    \item From the initial gain factors, calculate the required input power, $P_{\rm in}$, to achieve an average arm power of $P_{\rm avg}$ and the individual $x$- and $y$-arm powers, $P_{\rm X}$ and $P_{\rm Y}$, respectively:
    \begin{align}
    P_{\rm in} &= \frac{4}{G_{\rm PR} (G_{\rm X} + G_{\rm Y})} \, P_{\rm avg}
    \label{eq:Ppr} \\
    P_{\rm X}  &= \frac{2 G_{\rm X}}{G_{\rm X} + G_{\rm Y}} \, P_{\rm avg} 
    \label{eq:Px} \\
    P_{\rm Y}  &= \frac{2 G_{\rm Y}}{G_{\rm X} + G_{\rm Y}} \, P_{\rm avg}
    \label{eq:Py}
    \end{align}
    \item Rescale the mirror maps in the two arms to the updated incident power levels, $P_{\rm X}$ and $P_{\rm Y}$.
    \item Rerun the \textsc{Finesse} simulation to obtain updated cavity gain factors, $G_{\rm X}'$, $G_{\rm Y}'$, and $G_{\rm PR}'$.
    \item From the updated gain factors, recalculate required input power, $P_{\rm in}'$, and the individual arm powers, $P_{\rm X}'$ and $P_{\rm Y}'$ (using eqs.~\ref{eq:Ppr}-\ref{eq:Py}).
    \item Compare $P_{\rm X}'$ and $P_{\rm Y}'$ to the assumed incident powers used to generate the mirror maps, $P_{\rm X}$ and $P_{\rm Y}$. If the values of $P_{\rm X}'$ and $P_{\rm X}$ or $P_{\rm Y}'$ and $P_{\rm Y}$ differ by more than 1\%, return to step~5.
    \item Finally, compare $P_{\rm in}'$ to a chosen target input power, $P_{\rm target}$. If $P_{\rm in}'$ and $P_{\rm target}$ differ by more than 1\%, rescale the average arm power by half the difference as
    \begin{equation}
        P_{\rm avg}' = \frac{P_{\rm target} + P_{\rm in}'}{2 P_{\rm in}'} \, P_{\rm avg}
    \end{equation}
    and return to step~1.
\end{enumerate}
We find that in practice, with a well-chosen initial $P_{\rm avg}$ value, the above procedure converges to a self-consistent thermal-optical solution in only 2-3 iterations.

For computational efficiency, the squeezer subsystem's filter cavity and injection optics are not explicitly modeled, but only considered on an effective level. Following Ref.~\cite{FROSTI_PRL}, a squeezed field is injected from the antisymmetric port at an effective level chosen to account for the phase noise of the squeezer subsystem and all attenuation losses. Frequency dependency of the squeezing is then achieved by optimizing the squeezing angle of the injected squeezed field at each frequency bin, such that the quantum noise of the detector at that particular frequency is minimized. We inject an effective squeezing level of 9~dB which, in the absence of thermal distortions, yields the A+ squeezing target of approximately 7.4~dB of observed broadband quantum noise reduction.

Given a set of wavefront errors due to laser beam heating, our Monte Carlo approach simulates our practical inability to perfectly \textit{infer} those errors across the full mirror aperture, which limits the precision to which we can identify the optimal power settings of the wavefront actuators and the optimal beam positioning. In each trial, we randomly sample the laser beam's $x$- and $y$-centroid coordinates, the ring heater power setting, and the FROSTI power setting for each test mass. These parameters are drawn from independent Gaussian distributions centered on the parameters' true optimal values, corresponding to perfectly centered beams and the wavefront actuator powers which minimize the beam intensity profile-weighted RMS wavefront errors. The standard deviations of the parameters represent the level of inferential uncertainty in practically determining their optimal values. We assume the parameter standard deviations listed in table~\ref{tab:uncertainties}.

\begin{figure*}[htbp]
    \centering
    \includegraphics[width=1\linewidth]{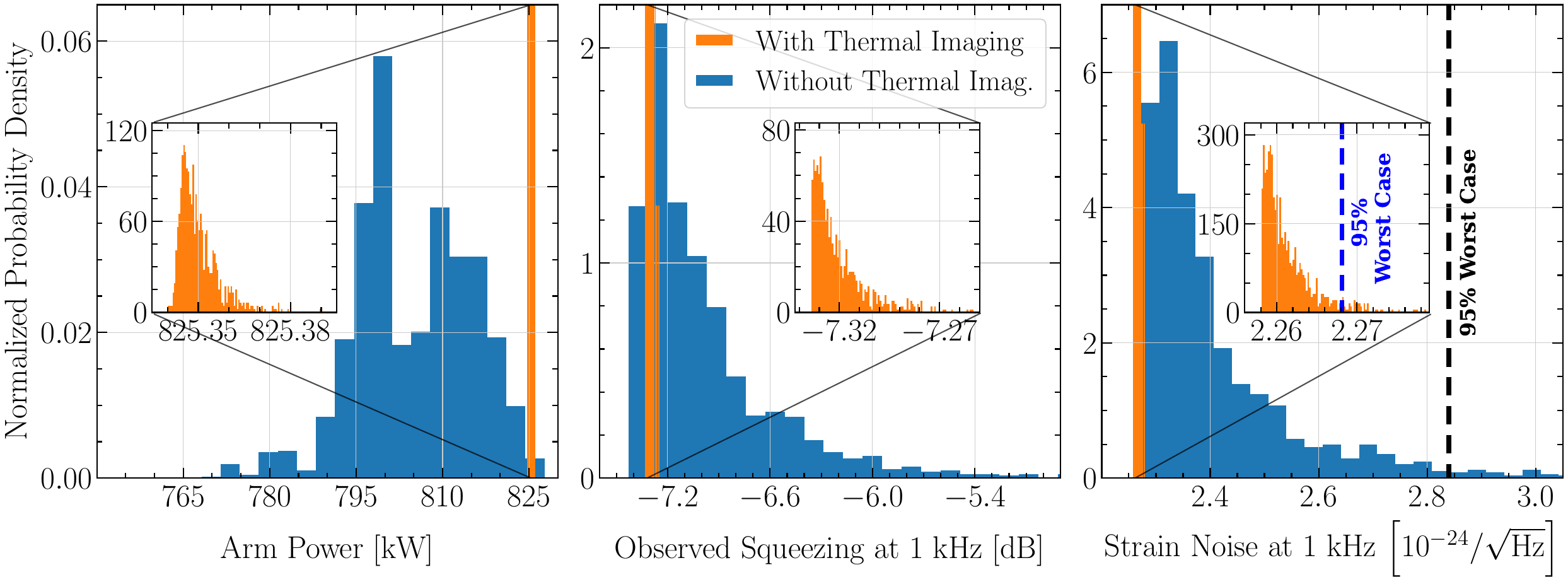}
    \caption{Distributions of the performance of a LIGO A+ detector with realistic uncertainties in the beam positions and the optimal power settings of the wavefront actuators. \textit{Left:} The achievable power in the arm cavities, with a fixed input power of 125~W. \textit{Middle:} The observed quantum squeezing level at 1~kHz, with a fixed effective input squeezing level of 9~dB. \textit{Right:} The resulting total quantum-noise-limited strain noise at 1~kHz. The dashed vertical lines indicate the worst possible outcomes at 95\% confidence.}
    \label{fig:interferometer_prediction}
\end{figure*}

Figure~\ref{fig:interferometer_prediction} shows the projected impact of the new wavefront sensing capabilities on the key performance metrics that determine LIGO's quantum-limited sensitivity. The two cases, with (orange) and without (blue) thermal imaging of the test masses, are obtained from randomized Monte Carlo trials assuming the inferential uncertainties in table~\ref{tab:uncertainties}. The left panel shows the probability distributions of laser power in the detector's arm cavities for a fixed input power of 125~W. The middle panel shows the probability distributions of observed squeezing at 1~kHz for a fixed effective injected squeezing level of 9~dB. The right panel shows the probability distributions of the total resulting strain noise at 1~kHz. Based on convergence testing of their 95th percentiles, we find that roughly 1000 trials are adequate to fully sample the distributions of the non-thermal imaging case, while only 180 trials are needed to fully sample the case with thermal imaging. In the non-thermal imaging case, we find that a small fraction of trials (2.7\%) result in exceptionally high contrast defect power which would be easily sensed and reduced via detector commissioning. To eliminate such cases, we veto any trials resulting in more than 100~mW of contrast defect power.

As is visible in the left two panels of figure~\ref{fig:interferometer_prediction}, there exist some cases in which trade-offs between optimizing the arm power (via the power recycling cavity coupling) and squeezing (via the signal recycling cavity coupling) lead to slightly better performance in one of these metrics, individually, without thermal imaging. However, the overall quantum noise level is consistently reduced in all cases when thermal imaging is employed, as shown in the right panel. We thus find that the higher precision inference enabled by test mass thermal imaging dramatically narrows the range of possible sensitivity outcomes, significantly improving the \textit{worst} possible outcome at 95\% confidence.

\begin{figure}[htbp]
    \centering
    \includegraphics[width=0.55\linewidth]{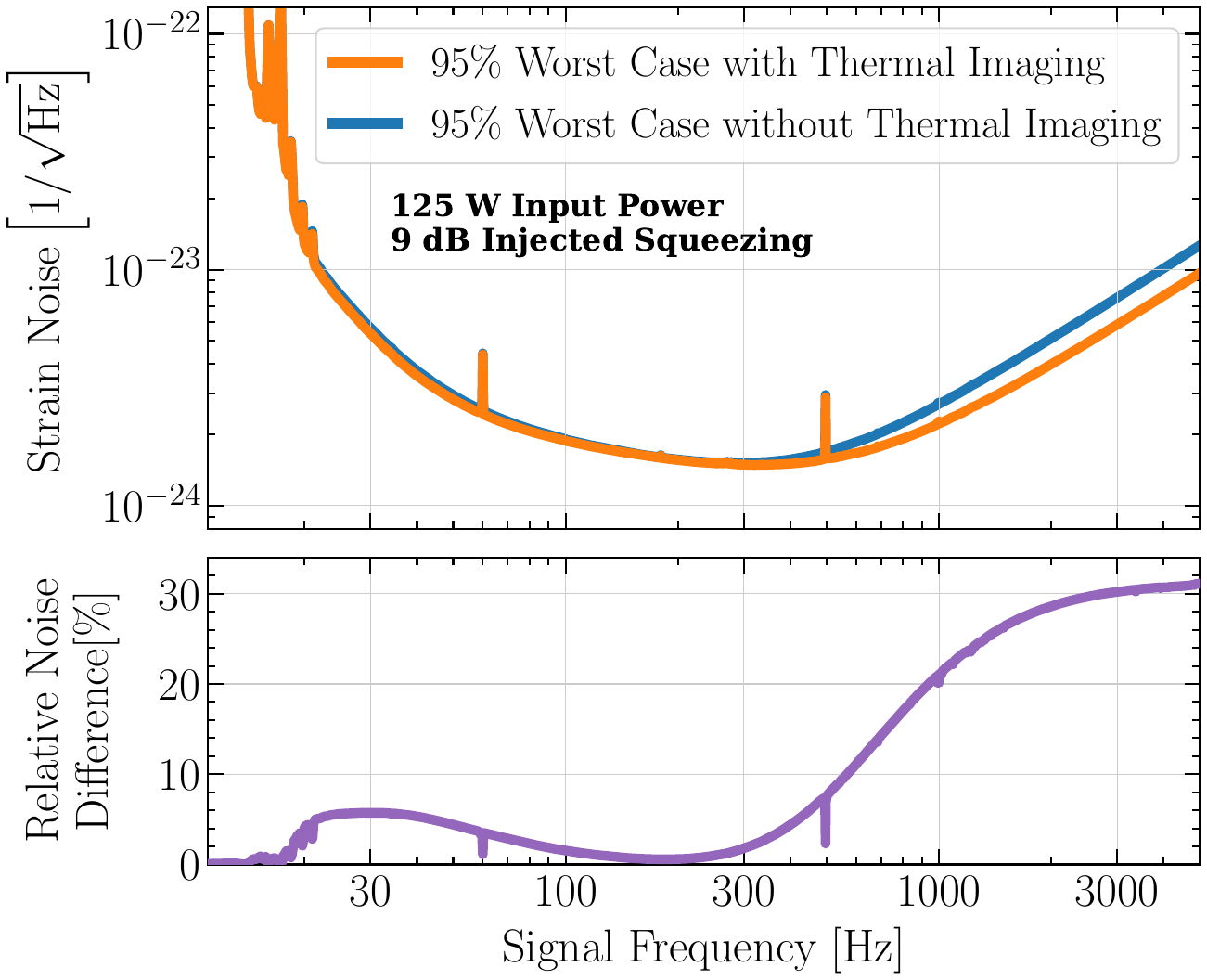}
    \caption{Comparison of the worst-case strain sensitivities of a LIGO A+ detector, at 95\% confidence, with and without thermal-imaging-based inference. These two cases correspond to the two cases denoted by the dashed vertical lines in the right panel of figure~\ref{fig:interferometer_prediction}.} 
    \label{fig:QN_prediction}
\end{figure}

This large difference in worst-case scenarios translates directly into a statistical improvement in the expected strain sensitivity of the detector, which is shown in figure~\ref{fig:QN_prediction} for the worst observed outcomes in 95\% of trials. That is, with 95\% confidence, the detector will achieve \textit{at least} the strain sensitivity indicated by these curves. Each case assumes the nominal A+ input power of 125~W and an effective injected level of 9~dB of frequency-dependent squeezing, as well as the nominal A+ thermal and technical noise contributions which are combined with the quantum noise curves calculated using \textsc{Finesse}. We find that thermal imaging of the test masses can enable up to a 31\% greater reduction of the noise floor in LIGO~A+, with maximum impact at the highest frequencies (which are quantum shot noise limited), corresponding to an increase of 10~Mpc in the sky-averaged detection range for binary neutron star mergers. The relative strain noise improvement diminishes towards lower frequencies because the A+ detectors become limited by other, non-quantum noise sources which are independent of the improved wavefront control.

\begin{figure*}[h]
    \centering
    \includegraphics[width=1\linewidth]{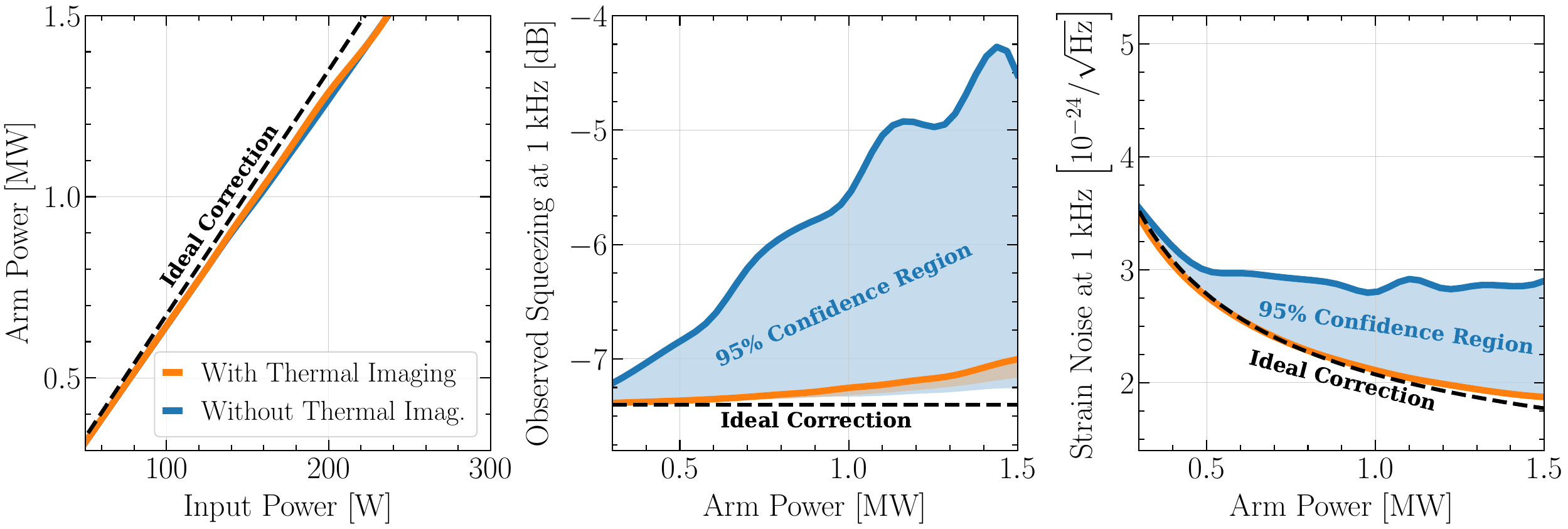}
    \caption{Projected performance of a LIGO A+ detector as the arm power is increased beyond the nominal target of 750~kW, toward the megawatt scale required for next-generation detectors. The shaded regions indicate the range of possible outcomes, at 95\% confidence, for the cases with and without thermal imaging. The dashed lines represent the case of perfect wavefront correction (no thermal distortions).}
    \label{fig:performance_95_arm_power}
\end{figure*}

Figure~\ref{fig:performance_95_arm_power} shows the LIGO detector's projected performance as we repeat the analysis of figure~\ref{fig:interferometer_prediction} while raising the laser power toward the higher levels required for LIGO $\rm A^{\#}$ and Cosmic Explorer. Each panel contains two shaded regions representing the range of outcomes spanned by the best 95\% of trials in each of the two cases, with (orange) and without (blue) test mass thermal imaging. The left panel compares the possible ranges of input power required to achieve a given arm power, at 95\% confidence. The middle and right panels compare the ranges of possible squeezing and strain sensitivity outcomes, for a given arm power level, at 95\% confidence. Without the enhanced inferential capabilities provided by thermal imaging, the probability of significant squeezing degradations grows rapidly above 750~kW due to the changing thermal state of the ITMs. With the new proposed capabilities, the detector's quantum noise performance is limited only by the intrinsic capability of the wavefront actuators to match the spatial profiles of the thermal distortions.

Finally, we note that Figure~\ref{fig:performance_95_arm_power} shows only the detector's performance for the main carrier field. In addition to the carrier, two sets of radio-frequency sidebands are injected at 9~MHz and 45~MHz for length control~\cite{Izumi:2017}. Wavefront distortions, particularly thermal lensing in the ITM substrates, affect the sideband fields differently than the carrier because the sidebands are non-resonant in the 4~km arm cavities. We include the 9~MHz and 45~MHz sidebands in our simulation and find that wavefront distortions are reduced sufficiently to maintain linear feedback control up to the most extreme carrier power level of 1.5~MW. However, further investigation of the effects on the sidebands assuming realistic sensing limits is needed and will be assessed in future work.

\section{Discussion}
\label{sec:conclusion}
Our analysis shows that real-time precision inference of the thermal state of the optics is vital to achieving greater quantum-limited sensitivity in gravitational-wave detectors. We have demonstrated that front-surface thermal imaging offers a promising emerging solution. The requirements for full-aperture wavefront correction are most stringent for the ITMs, due to wavefront distortions shifting the higher-order mode resonances of the signal recycling cavity~\cite{AsharpTCSReqs:2022}, and the sensing requirements identified in this paper can satisfy them. Our assumption of a thermal steady-state FEA model is a realistic assumption for future detectors with the introduction of the CHETA offline laser heating system (see section~\ref{sec:intro}). We anticipate that such an offline heating system will be extended to all four test masses by LIGO~$\rm A^{\#}$ and will be incorporated into the baseline Cosmic Explorer design. The impact of this work will thus extend far beyond A+, as it also delivers a key enabling sensing capability for next-generation gravitational-wave observatories.

Although we assume uniform coating absorption in this work, the presence of point absorbers, should they remain an issue for future detectors, can be handled through a straightforward generalization of our technique. As has already been demonstrated in Virgo, a point absorber will be easily visible in a thermal image of the test mass' front surface~\cite{VirgoLogbookPA}. Since the point absorber lies in a fixed position, its surface coordinates can be extracted from the thermal image and a point absorber heating source added to the FEA model of the test mass at that position. This requires introducing only one additional free parameter in the FEA model (per point absorber), representing the total amount of power absorbed by the point absorber. Due to the large difference in the spatial scales over which the absorption occurs, we expect there to be little degeneracy between the inferred powers of the point absorber and the other heating sources considered in this work.

Finally, the practical design of such a thermal imaging system will need to consider the detailed radiative environment surrounding each test mass (such as the suspension cage, reaction chain, and vacuum chamber walls), in order to evaluate its potential to bias the thermal state inference. These effects are represented by the $\delta P \, \hat{T}_{\rm \delta}(x, y)$ term in equation~\ref{eq:total_temp}. However, the successful demonstration of the proposed sensing technique during the in-vacuum testing of the FROSTI prototype suggests that such environmental effects are reasonably small~\cite{Tyler_2025_Optica}. This experiment found the average bulk temperature rise of the test mass to depend on the detailed radiative environment, but it found the \textit{relative} point-to-point temperature variation across the test mass' surface to be only weakly sensitive to this environment (relative to an idealized FEA model of the test mass). Moreover, current LIGO~$\rm A^{\#}$ and Cosmic Explorer planning incorporates low-emissivity gold shields around the barrel of each test mass. Their purpose is to reduce the amount of required thermal actuator power, but they will also serve to shield the optics from environmental heating sources. As a result, the impact of near-field sources in the surrounding suspension will likely be significantly reduced.

%
% Each of the commands below will create an unnumbered section with the appropriate heading.
% Remove any sections that are not relevant for your article.
% All sections except suppdata will be removed if the [anonymous] option is used.
% See iopjournal-guidelines.pdf for more information.
%

\data{Data underlying the results presented in this paper are not publicly available at this time but may be obtained from the authors upon reasonable request.}
% For more information on IOP Publishing's research data policy see: https://publishingsupport.iopscience.iop.org/questions/research-data/

\ack{The authors thank Georgia Mansell and the Cosmic Explorer mode sensing and control (MSC) design team for helpful comments during the preparation of this manuscript. This material is based upon work supported by the National Science Foundation (NSF) under Award Nos. PHY-2309006 and PHY-2409496. This paper carries LIGO Document Number LIGO-P2500333.}

\bibliographystyle{iopart-num}
\bibliography{references}

\end{document}